\documentclass{aastex631}

\shorttitle{AASTeX v6.3.1 Sample article}
\graphicspath{{./}{picture/}}  
\usepackage{amsmath}  
\usepackage{booktabs} 
\usepackage{url} 
\usepackage{amssymb}
\usepackage{rotating}
\usepackage{graphicx}
\begin{document}
	
	\title{A systematic search for physical associations between fast radio bursts and astrophysical transients}
	
	\author{Hao-Hao Chen}
	\author{Wen-Tao Xu}
	\author{Xin-Yu Liang}
	\author{Ming-Xuan Lu}
	\author{Can-Min Deng}
	\affiliation{Guangxi Key Laboratory for Relativistic Astrophysics, School of Physical Science and Technology, Guangxi University, Nanning 530004, People's Republic of China; dengcm@gxu.edu.cn}
	
	\begin{abstract}
	The physical origin of fast radio bursts (FRBs) remains an unsolved mystery in astrophysics, with the magnetar central engine model as the leading framework. Systematically searching for physical associations between FRBs and the energetic astrophysical transients (ATs) that form magnetars provides a critical test of this scenario, and key clues to FRB progenitors.
	We perform a systematic search for FRB-AT associations using a sample of 3765 unique FRBs, combining the second CHIME/FRB catalog with 124 additional localized FRBs with measured redshifts. We develop a 3D Bayesian inference framework that jointly incorporates angular separation, positional uncertainty, and redshift constraints to quantify the association probability of candidate pairs.
	Through spatial cross-matching, we identify 14 FRB-optical transient and 15 FRB-gamma-ray burst (GRB) candidate pairs. Our framework recovers the previously reported high-significance association between FRB 20180916B and AT 2020hur, with an association probability of 0.9998. For the proposed candidate FRB 20190309A and short GRB 060502B, our analysis yields an association probability of 0.83, which is insufficient to claim statistically significant association. No new statistically significant FRB-AT associations are found for all remaining candidates. Our work demonstrates that small angular separation alone is insufficient to confirm FRB-AT associations, and high-precision FRB localization is essential for definitive identification. 
	\end{abstract}
	
	\keywords{Fast Radio Bursts; Gamma Ray Bursts; Bayesian Inference}
	
	\section{Introduction}

Fast radio bursts (FRBs) are millisecond-duration flashes of radio emission with brightness temperature exceeding $10^{35}$ K \citep{Lorimer_2007,Thornton_2013}. Their large dispersion measures (DMs) firmly establish extragalactic distances, and interferometric localizations have identified host galaxies spanning from dwarf star-forming systems to massive spirals \citep{Chatterjee_2017, Tendulkar_2017, Fong_2021, Ravi_2023}. Over the past decade, thousands of FRBs have been discovered by CHIME/FRB \citep{CHIME_data2026}, revealing both apparently one-off events and highly active repeaters \citep{Spitler_2016,Fonseca_2020,2021ApJ...923....1P,2022ApJ...926..206Z}. Despite rapid observational progress, the physical nature of FRBs remains unresolved \citep{Zhang_2023}. Although the vast majority of FRBs are confirmed to be of extragalactic origin, the Galactic FRB 20200428D was found to be highly coincident in time with an X-ray burst from the soft gamma repeater SGR 1935+2154 \citep{CHIME_2020, Bochenek_2020,Ridnaia_2021,2021NatAs...5..414K,2021NatAs...5..408Y,2021NatAs...5..378L}. This observational fact demonstrates that magnetars can serve as the central engines for at least a fraction of FRBs.

Gamma-ray bursts (GRBs) are generally produced by either the core collapse of massive stars (long GRBs) or the mergers of binary compact stars (short GRBs) \citep{2018pgrb.book.....Z}. 
Similarly, other energetic astrophysical transients, such as core-collapse supernovae (CCSNe) and superluminous supernovae (SLSNe), mark the explosive deaths of massive stars \citep{2005NatPh...1..147W,2012Sci...337..927G}. Crucially, these catastrophic channels are highly likely to leave behind a massive, rapidly spinning, and highly magnetized neutron star—a magnetar. These newborn magnetars can act as the central engines of FRBs. 
Because the compact remnant can survive and remain active for an extended period, and the surrounding dense ejecta needs time to expand and become transparent to low- frequency radio waves, the FRB emission is theoretically expected to occur with a time delay ranging from months to decades \citep{Margalit_2018,Wang_2020}. Therefore, we can naturally expect a spatial overlap between the ATs and the locations of FRBs.

Driven by this theoretical expectation, researchers have been dedicated to searching for spatial associations between FRBs and historical GRBs or supernovae. On the one hand, targeted radio observations of historical catastrophic remnants have primarily yielded non-detections \citep{2019ApJ...886...24L, 2019ApJ...887..252M, Palliyaguru_2020}. On the other hand, systematic searching analyses using archival catalogs have successfully identified several promising association candidates. 
For instance, \citet{Wang_2020} and \citet{Lu_2024} identified spatial coincidences for FRB 20171209–GRB 110715A and FRB 190309A–GRB 060502B, respectively. Both pairs were proposed as physically self-consistent candidates, yet they possess only marginal statistical significance. Similarly, cross-matching efforts using the first CHIME/FRB catalog identified FRB 20190412B and SN 2009gi as another compelling pair. While their spatial overlap might simply be a chance occurrence, this pair satisfies rigorous physical constraints \citep{2025ApJ...992..127L}. In contrast to these cases, \citet{Li_2022} reported a highly probable association between the repeating FRB 20180916B and the optical transient AT2020hur, which exhibits a low chance coincidence probability.

The recent release of the second CHIME/FRB catalog (comprising 4,539 FRBs) \citep{CHIME_data2026}, supplemented by 124 localized FRBs with measured redshifts outside of this catalog \citep{2025arXiv251102155T}, provides an unprecedented opportunity to conduct a more comprehensive search. In this paper, we develop a 3D Bayesian framework on this expanded dataset to systematically search for potential physical associations between FRBs and various ATs. The structure of this paper is as follows: Section \ref{Data} describes the data samples of FRBs and ATs used in our analysis. Section \ref{Bayesian} details the Bayesian framework. The results of our systematic search and the discussion are presented in Section \ref{Result} and \ref{Discussion}, followed by our main conclusions in Section \ref{conclusion}.

\section{The Data}\label{Data}
\subsection{Sample}
To systematically search for physical associations between FRBs and ATs, we construct our sample using comprehensive public catalogs available for FRBs, GRBs \footnote{\url{https://www.mpe.mpg.de/~jcg/grbgen.html}}, 
optical transients \footnote{\url{https://www.rochesterastronomy.org/supernova.html}}(OTs, specifically supernovae) and X-ray transients (XTs) \citep{CHIME_data2026, 2025arXiv251102155T, Guo_2025}. For each event, we primarily extract the spatial coordinates (Right Ascension and Declination), their localization errors, and the redshift, which are the essential inputs for our 3D Bayesian framework.

\textbf{FRB Sample:} Our primary FRB dataset is constructed from two sources: the recently released second CHIME/FRB catalog \citep{CHIME_data2026} with 124 additional localized FRBs from \citet{2025arXiv251102155T}. We carefully clean the combined dataset by excluding multiple bursts from known repeating FRBs (retaining only one event for a repeater) and removing the same sources that appear in both catalogs. Consequently, we establish a final sample of 3,765 FRBs for our cross-matching analysis.

\textbf{AT Sample:} For the GRB dataset, we select a total of 701 GRBs from the catalog that possess well-constrained localized positions and measured redshifts, which are further classified into 626 long GRBs (LGRBs) and 75 short GRBs (SGRBs). 
The optical source sample, which primarily consists of supernovae and other optical transients, is collected from the Latest Supernovae database. Similarly, we filter this database to include only those events that possess definitive localization and explicitly measured redshifts (either from the transient itself or its host galaxy). This selection process yields a final sample of 51,054 optical sources. Furthermore, we supplement our dataset by incorporating 17 X-ray transients (XTs) extracted from the compilation provided by \citet{Guo_2025}.

\subsection{Redshift Distribution} \label{p_z}
To evaluate the posterior probability using our Bayesian framework,  we need to construct the redshift probability density functions for both samples. For the AT sample, we apply Kernel Density Estimation (KDE) independently to the 701 GRBs and the 51,054 optical sources, thereby deriving redshift probability density $p_\mathrm{AT}(z)$ for each transient type.

For the FRBs, we infer their individual redshift probability density from their extragalactic dispersion measures (${\rm DM_E}$). The ${\rm DM_E}$ is determined by subtracting the Milky Way interstellar medium (${\rm DM_{MW}}$) and galactic halo (${\rm DM_{halo}}$) contributions from the observed DM:
\begin{equation}\label{eq:DM_E}
	{\rm DM_E}\equiv {\rm DM_{obs}}-{\rm DM_{MW}}-{\rm DM_{halo}}={\rm DM_{IGM}}+\frac{{\rm DM_{host}}}{1+z}.
\end{equation}
We fix ${\rm DM_{halo}} = 50~{\rm pc~cm^{-3}}$ and use the NE2001 model to estimate ${\rm DM_{MW}}$ \citep{2004ASPC..317..211C}. The mean IGM contribution is calculated assuming the standard $\Lambda$CDM model with Planck 2018 cosmological parameters \citep{2014ApJ...783L..35D, 2020Natur.581..391M, 2020A&A...641A...6P}:
\begin{equation}\label{eq:DM_IGM}
	\langle{\rm DM_{IGM}}(z)\rangle=\frac{21cH_\mathrm{0}\Omega_\mathrm{b}f_{\rm IGM}}{64\pi Gm_\mathrm{p}}\int_0^z\frac{1+z'}{\sqrt{\Omega_\mathrm{m}(1+z')^3+\Omega_\mathrm{\Lambda}}}dz'.
\end{equation}
Theoretical simulations show that the probability distribution for ${\rm DM_{IGM}}$ has a flat tail at large values, which can be fitted with the following function \citep{2020Natur.581..391M,2021ApJ...906...49Z}:
\begin{equation}
	p_{\rm IGM}(\Delta)=A\Delta^{-\beta}\exp\left[-\frac{(\Delta^{-\alpha}-C_0)^2}{2\alpha^2\sigma_{\rm IGM}^2}\right], ~~~\Delta>0,
\end{equation}
where $\Delta\equiv{\rm DM_{IGM}}/\langle{\rm DM_{IGM}}\rangle$,  $\sigma_{\rm IGM}=Fz^{-1/2}$ ($F$ is a free parameter) is the effective standard deviation, $\alpha$ and $\beta$ are related to the inner density profile of gas in haloes (fixed $\alpha=\beta=3$), $A$ is a normalization constant, and $C_0$ is chosen such that the mean of this distribution is unity. Simultaneously, the host galaxy contribution ${\rm DM_{host}}$ is assumed to follow a log-normal distribution \citep{2020Natur.581..391M,2021ApJ...906...49Z}:

\begin{eqnarray}\nonumber\label{eq:P_host}
	p_{\rm host}({\rm DM_{host}}|\mu_\mathrm{host},\sigma_{\rm host})&=&\frac{1}{\sqrt{2\pi}{\rm DM_{host}}\sigma_{\rm host}}\\
	&\times&\exp\left[-\frac{(\ln {\rm DM_{host}}-\mu_\mathrm{host})^2}{2\sigma_{\rm host}^2}\right],
\end{eqnarray}
where $\mu_\mathrm{host}$ is the mean value and $\sigma_{\rm host}$ is the standard deviation of $\ln {\rm DM_{host}}$.

\begin{figure}[ht!]
	\centering
	\includegraphics[width=\columnwidth]{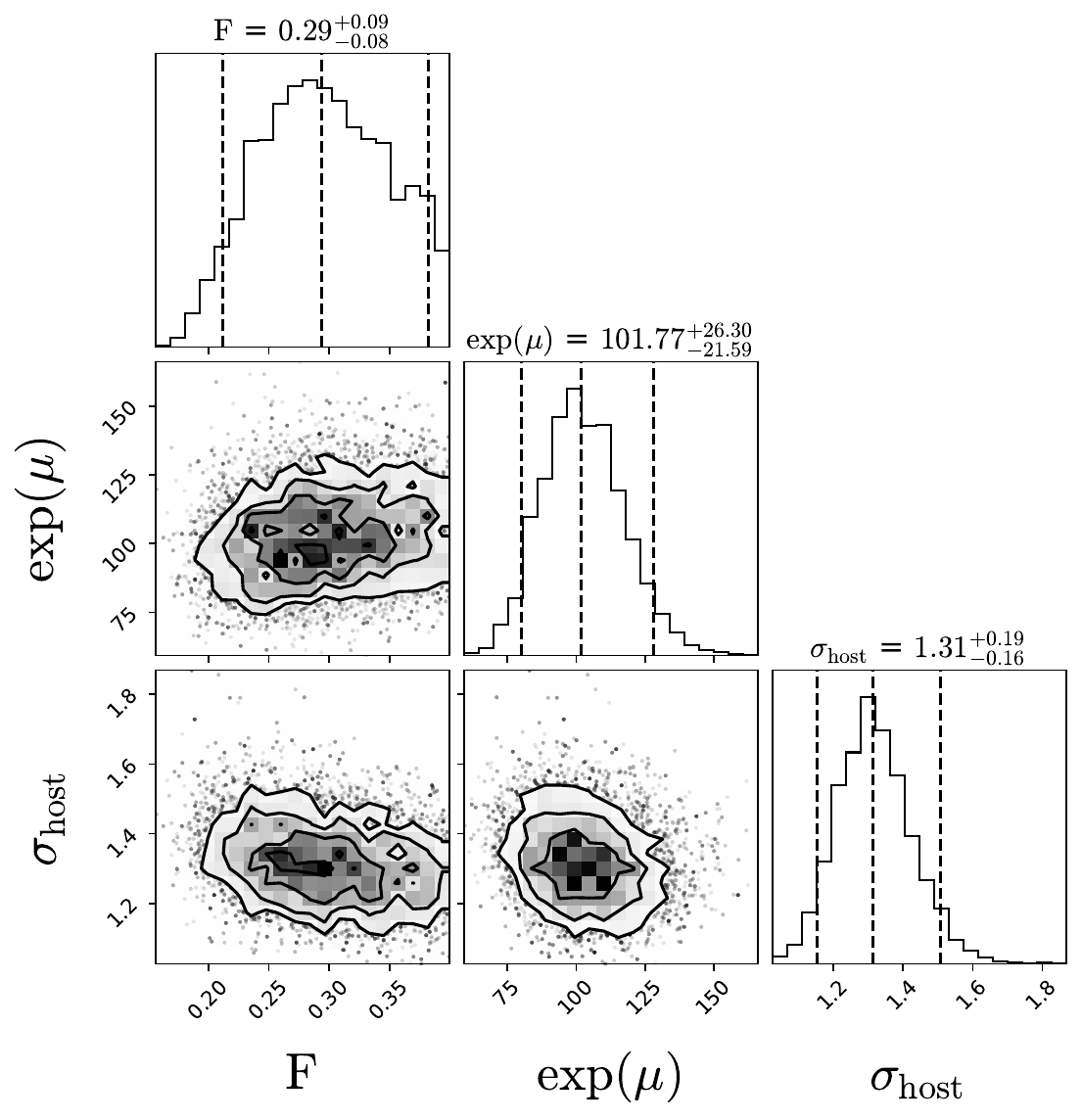}
	\caption{The posterior distributions of the parameters for the 133 localized FRBs. The updated parameters are $F = 0.29_{-0.09}^{+0.08}$, $e^\mu = 101.77_{-26.30}^{+21.59}$, and $\sigma_{\rm host} = 1.31_{-0.16}^{+0.19}$ (with uncertainties reported at the 90\% credible interval).}
	\label{fig:frb_parameters} 
\end{figure}

Given these assumptions, the probability distribution of observing a specific ${\rm DM_E}$ at redshift $z$ is the convolution of the IGM and host galaxy distributions:
\begin{equation}\label{eq:P_E}
	p({\rm DM_E}|z)=\int_0^{(1+z)\rm DM_E}p_{\rm host}({\rm DM_{host}}|\mu_\mathrm{host},\sigma_{\rm host}) p_{\rm IGM}\left({\rm DM_E}-\frac{\rm DM_{host}}{1+z}\Big|F,z\right)d{\rm DM_{host}}.
\end{equation}

The likelihood that we observe a sample of FRBs with ${\rm DM_{E,\it i}}$ at redshift $z_i$ is given by
\begin{equation}
	\mathcal{L}({\rm FRBs}|F,\mu_\mathrm{host},\sigma_{\rm host})=\prod_{i=1}^Np_{i}({\rm DM_{E,\it i}}|z_i),
\end{equation}
where $N$ is the total number of FRBs. Given the FRB data ($z_i,{\rm DM_{E,\it i}}$), the posterior probability distribution of the parameters ($F,\mu_\mathrm{host},\sigma_{\rm host}$) is given according to Bayes theorem by
\begin{equation}
	P(F,\mu_\mathrm{host},\sigma_{\rm host}|{\rm FRBs})\propto\mathcal{L}({\rm FRBs}|F,\mu_\mathrm{host},\sigma_{\rm host})P_0(F,\mu_\mathrm{host},\sigma_{\rm host}),
\end{equation}
We use the 133 localized FRBs with definitively measured redshifts to perform a Bayesian parameter inference \citep{2025arXiv251102155T}. Finally, we update the parameters to $F = 0.29_{-0.09}^{+0.08}, \quad e^{\mu_\mathrm{host}} = 101.77_{-26.30}^{+21.59}, \quad \sigma_{\rm host} = 1.31_{-0.16}^{+0.19}.$ (with uncertainties reported at the 90\% credible interval), as shown in Figure \ref{fig:frb_parameters}.

According to Bayes' theorem, the posterior probability density function of the redshift for a given FRB, denoted as $p(z|{\rm DM_E})$, is proportional to the product of the likelihood $p({\rm DM_E}|z)$ and the prior redshift distribution $p_{\rm prior}(z)$. Although the true FRB volumetric rate likely traces the cosmic star formation history or stellar mass, we adopt, for simplicity, a uniform prior over the range $[0, 10]$, which then becomes a constant and drops out of the normalization. Consequently, the final normalized posterior probability density function simplifies directly to the normalized likelihood:
\begin{equation}
	p(z|\rm DM_E) = \frac{p({\rm DM_E}|z)}{\int_0^{10} p({\rm DM_E}|z) dz},
\end{equation}
which serves as the direct redshift input for our 3D Bayesian framework.

\section{The method: Bayesian framework for Association Probability}\label{Bayesian}

We consider two mutually exclusive hypotheses:

$H_{1}$: Physical association: the FRB and the AT share the same true position and redshift.

$H_{0}$: Chance alignment: the AT is a background object, with its position and redshift independently drawn from the background distribution.

\subsection{Prior}
The prior probabilities are set as $P(H_{1})=\pi_{1}$ and $P(H_{0})=1-\pi_{1}$, where $\pi_{1}$ is a small value reflecting the rarity of true associations.
A conservative value is $\pi_{1}=1/(N_\mathrm{FRB}N_\mathrm{AT})$, which is effective for suppressing false positives driven by the look-elsewhere effect in large-scale catalog cross-matching.  In this work, we adopt this conservative value.

\subsection{Likelihood Functions}

We condition on the FRB data and derive the conditional probability densities of observing the AT data under each hypothesis.

\subsubsection{Likelihood under \texorpdfstring{$H_{1}$}{H1}}

Under $H_{1}$, the true positions coincide. The observed angular separation arises solely from the FRB positional error.
In the tangent-plane approximation, the measurement error is a bivariate Gaussian, so the radial distance $\theta$ follows a Rayleigh distribution \citep{2021ApJ...911...95A}:

\begin{equation}
	p(\theta|H_{1})=\frac{\theta}{\sigma^{2}}\exp\left(-\frac{\theta^{2}}{2\sigma^{2}}\right) \quad (\theta\ge0).
\end{equation}
This is derived by transforming Cartesian coordinates $(x,y)$ \footnote{Given by $p(x,y) = (2\pi\sigma^{2})^{-1}\exp[-(x^{2}+y^{2})/(2\sigma^{2})]$.} into polar coordinates $(\theta,\phi)$ and integrating over $\phi$. And the $\sigma$ represents the effective combined positional uncertainty of the FRB-AT candidate pair. Assuming that the measurement errors in Right Ascension (RA) and Declination (Dec) are independent, we calculate the equivalent radial positional error for each source. To accurately reflect the true angular scale on the celestial sphere, we apply the standard spherical projection correction to the RA uncertainty \citep{2017A&A...597A..89P}:
\begin{equation}
	\begin{aligned}
		\sigma_{\mathrm{FRB}}=\sqrt{\frac{(\sigma_{\mathrm{FRB,RA}}\cos\mathrm{Dec}_{\mathrm{FRB}})^2+\sigma_{\mathrm{FRB,Dec}}^2}{2}}, \\
		\sigma_{\mathrm{AT}}=\sqrt{\frac{(\sigma_{\mathrm{AT,RA}}\cos\mathrm{Dec}_{\mathrm{AT}})^2+\sigma_{\mathrm{AT,Dec}}^2}{2}}.
	\end{aligned}
\end{equation}
Since the positional measurements of the FRB and the AT are entirely independent, the total effective standard deviation $\sigma$ is given by the quadrature sum of their respective uncertainties $\sigma = \sqrt{\sigma_{\mathrm{FRB}}^2 + \sigma_{\mathrm{AT}}^2}$ \citep{2008ApJ...679..301B, 2021ApJ...911...95A}.

Because the true redshifts are identical under $H_{1}$, the probability density of observing the AT redshift $z_{\rm AT}$ is simply the FRB redshift probability density function evaluated at that specific redshift:
\begin{equation}
	p(z_{\rm AT}|H_{1})=p(z=z_{\rm AT}|{\rm DM_E}).
\end{equation}
Assuming independence between position and redshift, the joint likelihood is the product:
\begin{equation}
	p(\theta,z_{\rm AT}|H_{1})=\frac{\theta}{\sigma^{2}}\exp\left(-\frac{\theta^{2}}{2\sigma^{2}}\right)p(z=z_{\rm AT}|{\rm DM_E}).
\end{equation}

\subsubsection{Likelihood under \texorpdfstring{$H_{0}$}{H0}}

Under $H_{0}$, the AT is a random background object.
Its position is uniformly distributed over the sky, so the probability density of finding it at angular separation $\theta$ from the FRB position (in a ring of width $d\theta$) is:
\begin{equation}
	p(\theta|H_{0})=\frac{2\pi \sin\theta}{A}\simeq\frac{2\pi\theta}{A},
\end{equation}
where $A$ is the sky coverage area. Its redshift follows the sample distribution $p_\mathrm{AT}(z)$:
\begin{equation}
	p(z_\mathrm{AT}|H_{0})=p_\mathrm{AT}(z=z_\mathrm{AT}),
\end{equation}
Again, position and redshift are independent, yielding the joint likelihood:
\begin{equation}
	p(\theta,z_\mathrm{AT}|H_{0})=\frac{2\pi\theta}{A}p_\mathrm{AT}(z=z_\mathrm{AT}).
\end{equation}

\subsection{Posterior Probability}
The Bayes factor $B$, which measures the relative support for $H_{1}$ versus $H_{0}$, is the ratio of the two likelihoods:
\begin{equation}
	B = \frac{p(\theta, z_\mathrm{AT} \mid H_1)}{p(\theta, z_\mathrm{AT} \mid H_0)}
	= \frac{ \dfrac{\theta}{\sigma^2} e^{-\theta^2/(2\sigma^2)} p(z_\mathrm{AT}|{\rm DM_E}) }{ \dfrac{2\pi\theta}{A} p_{\mathrm{AT}}(z_\mathrm{AT}) }
	= \frac{A}{2\pi\sigma^2} \cdot \frac{p(z_\mathrm{AT}|{\rm DM_E})}{p_{\mathrm{AT}}(z_\mathrm{AT})} \cdot \exp\left(-\frac{\theta^2}{2\sigma^2}\right), \label{eq:B}
\end{equation}
where $p(z_\mathrm{AT}|{\rm DM_E}) \equiv p(z=z_\mathrm{AT}|{\rm DM_E}) $ and ${p_{\mathrm{AT}}(z_\mathrm{AT})}\equiv {p_{\mathrm{AT}}(z=z_\mathrm{AT})}$.

Applying Bayes' theorem \citep{Sivia_2006, Gregory_book},  the posterior probability of $H_{1}$ given the data is:
\begin{equation}
	P(H_{1}|\theta,z_\mathrm{AT}) = \frac{\pi_{1}B}{\pi_{1}B+(1-\pi_{1})} \simeq \frac{\pi_{1}B}{\pi_{1}B+1},
\end{equation}
given that the prior probability $\pi_{1}$ is a very small number ($\pi_{1} \ll 1$, which implies $1-\pi_{1} \simeq 1$).
Substituting the Bayes factor and rearranging, we obtain the final expression:
\begin{equation} \label{P_asso}
	\begin{aligned}
		P_\mathrm{asso} = P(H_{1}|\theta,z_\mathrm{AT})= \left[ 1 + N_\mathrm{FRB}  \cdot {2\pi\sigma^{2}}\lambda_\mathrm{AT} \cdot \frac{p_{\mathrm{AT}}(z_\mathrm{AT})}{p(z_\mathrm{AT}|{\rm DM_E})} \cdot \exp\left(\frac{\theta^{2}}{2\sigma^{2}}\right) \right]^{-1},
	\end{aligned}
\end{equation}
where $P_\mathrm{asso}$ denotes the association probability, $\lambda_\mathrm{AT}=N_\mathrm{AT}/A$ is the surface number density of ATs. Based on the derivation above, we can compute the association probability for any given pair of candidates. When redshift information of the AT is unavailable, the factors related to redshift in this formula, $p_{\mathrm{AT}}(z_\mathrm{AT})/p(z_\mathrm{AT}|{\rm DM_E})$, can be removed, enabling computation of the position association probability solely.

{Here, our formalism does not incorporate temporal correlation between FRBs and their potential associated transients. This is because the physical origin of FRBs remains unresolved, and any such temporal correlation is inherently model-dependent. The current mainstream theory posits that FRBs originate from magnetars, which can be formed via two primary channels: the core collapse of massive stars, and the merger of binary compact objects \cite{2013ApJ...771L..26G, Margalit_2019, 2026ApJ...999..102H}. Both formation channels can produce associated astrophysical transients: the former can power long-duration gamma-ray bursts (GRBs)\cite{1992ApJ...392L...9D, 1998PhRvL..81.4301D, 10.1111/j.1365-2966.2011.18280.x} and core-collapse supernovae \cite{Yu_2017, 10.1093/mnrasl/slaf114}, while the latter can produce short-duration GRBs and merger-related optical transients. Under the core-collapse formation scenario, FRB emission is expected to occur up to thousands of years after magnetar formation, meaning FRB events should postdate the associated GRBs, supernovae and other cosmic transients \cite{Margalit_2018, Wang_2020}. However, it cannot be ruled out that a fraction of FRBs are produced via alternative channels. For example, FRBs may be generated during the inspiral and merger of binary neutron stars or white dwarf systems \cite{2013ApJ...776L..39K}; in this case, FRB emission is expected to occur earlier than the associated transients such as GRBs and supernovae. Therefore, given the lack of a robust, model-independent consensus on the temporal correlation between FRBs and astrophysical transients, we do not consider temporal constraints in our analysis.}

	\section{Results} \label{Result}
	We first carried out spatial cross-matching between the FRB and astrophysical transient (AT) datasets, with selection criteria tailored to the distinct sky number densities of different transient classes. Given the extremely high number surface density of optical transients (OTs), we computed the angular separation $\theta$ for all FRB-OT pairs and retained only those with $\theta < 1$ arcmin as initial candidates. For the significantly sparser FRB-GRB and FRB-XT samples, we instead adopted a conservative selection rule, requiring candidate counterparts to fall within the positional error regions of the corresponding FRBs. This initial spatial screening yielded a total of 14 FRB-OT candidate pairs and 15 FRB-GRB candidate pairs (the latter including one short GRB).
	To quantify the statistical significance of physical association for these spatially coincident candidates, we applied our 3D Bayesian framework to compute the posterior association probability ($P_{\rm asso}$) for each individual pair. The full observational properties of all candidates and the corresponding calculation results are presented in Table \ref{tab:frb_sn} (for FRB-OT pairs) and Table \ref{tab:frb_grb} (for FRB-GRB pairs).
	
	For the 14 FRB-OT pairs listed in Table \ref{tab:frb_sn}, the FRB 20180916B-AT 2020hur pair yields an association probability of 0.9998. We note that AT 2020hur has no direct redshift measurement, so this probability is calculated solely based on the celestial positional information of the two sources, without incorporating redshift constraints. The remaining 13 FRB-OT pairs all have posterior probabilities lower than $10^{-3}$ under our adopted prior, among which three pairs associated with FRB 20250316A have posterior probabilities approaching zero. For the 15 FRB-GRB pairs listed in Table \ref{tab:frb_grb}, all probabilities are calculated with our Bayesian framework incorporating both angular separation and redshift constraints. The FRB 20190309A-GRB 060502B pair has an association probability of 0.83, two other pairs have marginal association probabilities below 0.11, and the remaining 12 pairs have association probabilities approaching zero. Overall, our calculations recover two previously reported association candidates\citet{Li_2022,Lu_2024}, with no new statistically significant FRB-AT association pairs identified in this work.

	\clearpage	
\begin{sidewaystable*}
	\centering
	\caption{\small{Properties and Association Probabilities of the 14 FRB-OT Candidate Pairs.}}
	\label{tab:frb_sn}
	\setlength{\tabcolsep}{2pt}
	
	\begin{tabular}{cccccccccccc} 
		\hline\hline 
		FRBs & ${\rm RA_{FRB}}$ & ${\rm Dec_{FRB}}$ & ${\rm DM_E}$ & OT & ${\rm RA_{OT}}$ & ${\rm Dec_{OT}}$ & $\theta$ & $z_{\rm OT}$ & Bayes Factor & $\sigma$ & $P_{\rm asso}$ \\
		& ( deg ) & (  deg ) & ( ${\rm pc\ cm^{-3}}$ ) & & (  deg ) & (  deg ) & ( deg ) & & & ( deg ) & \\
		\hline
		FRB 20180916B & $29.5031258$ & $65.716754$ & $101.0481$ & AT2020hur$^{1}$ & $29.503125$ & $65.71675$ & $ 4.23 \times 10^{-6}$ & $0.0337$ & $\sim 10^{10}$ & $ 2.78 \times 10^{-4}$ & 99.98\% \\
		FRB 20221025G & $13.286$ & $53.995$ & $1046.1817$ & SN2024adfq & $13.310$ & $53.994$ & $0.014$ & $0.2285$ & $\sim 10^{5}$ & $0.082$ & $\sim 10^{-3}$ \\
		FRB 20220224A & $43.438$ & $13.006$ & $254.9588$ & SN2025pfh$^{1}$ & $43.426$ & $13.016$ & $0.015$ & $0.012142$ & $\sim 10^{5}$ & $0.077$ & $\sim 10^{-3}$ \\
		FRB 20230614C & $245.838$ & $37.266$ & $109.2904$ & PS15boe$^{1}$ & $245.841$ & $37.253$ & $0.013$ & $0.033843$ & $\sim 10^{5}$ & $0.209$ & $\sim 10^{-4}$ \\
		FRB 20220424D & $188.675$ & $62.330$ & $3117.1339$ & AT2025mbm$^{1}$ & $188.672$ & $62.333$ & $0.003$ & $0.135823$ & $\sim 10^{4}$ & $0.112$ & $\sim 10^{-4}$ \\
		FRB 20190317D & $238.092$ & $47.049$ & $1034.0295$ & AT2019pmq$^{1}$ & $238.100$ & $47.061$ & $0.013$ & $0.162$ & $\sim 10^{4}$ & $0.138$ & $\sim 10^{-4}$ \\
		FRB 20221019B & $302.684$ & $86.489$ & $1277.2989$ & SN2022foj & $302.747$ & $86.493$ & $0.006$ & $0.058$ & $\sim 10^{4}$ & $0.038$ & $\sim 10^{-5}$ \\
		FRB 20190502A & $164.958$ & $59.947$ & $540.8446$ & SN2020adii & $164.973$ & $59.954$ & $0.010$ & $0.045077$ & $\sim 10^{4}$ & $0.102$ & $\sim 10^{-5}$ \\
		FRB 20200813C & $286.265$ & $34.390$ & $318.2156$ & SN2021soe & $286.251$ & $34.392$ & $0.012$ & $0.012$ & $\sim 10^{4}$ & $0.219$ & $\sim 10^{-5}$ \\
		FRB 20210408A & $22.547$ & $19.157$ & $431.5286$ & AT2020acvc$^{1}$ & $22.536$ & $19.169$ & $0.016$ & $0.045438$ & $\sim 10^{4}$ & $0.131$ & $\sim 10^{-5}$ \\
		FRB 20210604B & $164.883$ & $47.487$ & $559.4512$ & AT2023jk$^{1}$ & $164.898$ & $47.499$ & $0.016$ & $0.100061$ & $\sim 10^{4}$ & $0.172$ & $\sim 10^{-5}$ \\
		FRB 20250316A & $182.435$ & $58.849$ & $79.52$ & SN2008X & $182.451$ & $58.850$ & $0.008$ & $0.0063$ & $\sim 0$ & $3.55 \times 10^{-4}$ & $\sim 0$ \\
		FRB 20250316A & $182.435$ & $58.849$ & $79.52$ & SN2009E & $182.457$ & $58.847$ & $0.012$ & $0.0063$ & $\sim 0$ & $3.55 \times 10^{-4}$ & $\sim 0$ \\
		FRB 20250316A & $182.435$ & $58.849$ & $79.52$ & AT2025erx$^{1}$ & $182.461$ & $58.841$ & $0.016$ & $0.006354$ & $\sim 0$ & $3.55 \times 10^{-4}$ & $\sim 0$ \\
		\hline
		\multicolumn{12}{l}{%
			\begin{minipage}{\linewidth}
				\smallskip
				\footnotesize $^{1}$ For these sources lacking direct redshift measurements, the redshifts of their host galaxies are used in the calculations.\\
				\textbf{Note.} Because AT2020hur lacks direct redshift measurements and is located within the same host galaxy as FRB 20180916B, we perform the Bayesian inference for this pair based solely on spatial positions without considering redshift constraints. In these calculations, the positional uncertainty of the OTs are set to 1 arcsec.
			\end{minipage}%
		} \\
	\end{tabular}
\end{sidewaystable*}
	\clearpage	
	
	\begin{sidewaystable*}
		\centering
		\caption{\small{Properties and Association Probabilities of the 15 FRB-GRB Candidate Pairs.}}
		\label{tab:frb_grb}
		
		\setlength{\tabcolsep}{2pt}
		
		\begin{tabular}{cccccccccccc} 
			\hline\hline 
			FRBs & ${\rm RA_{FRB}}$ & ${\rm Dec_{FRB}}$ & ${\rm DM_E}$ & GRB & ${\rm RA_{GRB}}$ & ${\rm Dec_{GRB}}$ & $\theta$ & $z_{\rm GRB}$ & Bayes Factor & $\sigma$ & $P_{\rm asso}$ \\
			& ( deg  ) & ( deg  ) & ( ${\rm pc\ cm^{-3}}$ ) & & ( deg  ) & ( deg  ) & ( deg  ) & & & ( deg  ) & \\
			\hline
			FRB 20190309A & $278.947$ & $52.407$ & $248.2137$ & SGRB 060502B & $278.938$ & $52.633$ & $0.226$ & $0.287$ & $ \sim 10^{6}$ & $0.204$ & $83.25\%$ \\
			FRB 20210917A & $184.026$ & $20.287$ & $1005.4806$ & GRB 140318A & $184.054$ & $20.200$ & $0.091$ & $1.02$ & $\sim 10^{5}$ & $0.256$ & $10.70\%$ \\
			FRB 20200721E & $69.174$ & $85.338$ & $530.3554$ & GRB 211023A & $72.300$ & $85.300$ & $0.258$ & $0.39$ & $\sim 10^{5}$ & $0.213$ & $9.92\%$ \\
			FRB 20191103B & $234.230$ & $16.145$ & $806.8435$ & GRB 150518A & $234.208$ & $16.267$ & $0.123$ & $0.256$ & $\sim 10^{4}$ & $1.021$ & $\sim 10^{-3}$ \\
			FRB 20200314H & $107.061$ & $27.079$ & $981.6112$ & GRB 160228A & $107.321$ & $26.950$ & $0.265$ & $1.64$ & $\sim 10^{3} $ & $0.246$ & $\sim 10^{-3}$ \\
			FRB 20181022F & $179.227$ & $17.111$ & $647.5824$ & GRB 250321D & $179.263$ & $17.383$ & $0.275$ & $4.368$ & $\sim 0 $ & $0.251$ & $\sim 0 $ \\
			FRB 20190429A & $281.038$ & $59.424$ & $363.2501$ & GRB 140713A & $281.113$ & $59.617$ & $0.197$ & $0.935$ & $\sim 0 $ & $0.198$ & $\sim 0 $ \\
			FRB 20200705D & $187.378$ & $47.879$ & $289.0007$ & GRB 250920B & $187.338$ & $47.850$ & $0.040$ & $2.2$ & $\sim 0 $ & $0.171$ & $\sim 0 $ \\
			FRB 20201026F & $14.285$ & $-5.724$ & $734.3873$ & GRB 251002A & $13.904$ & $-5.550$ & $0.417$ & $2.178$ & $\sim 0 $ & $0.349$ & $\sim 0 $ \\
			FRB 20210504C & $238.095$ & $78.505$ & $625.6949$ & GRB 060510B & $239.142$ & $78.567$ & $0.217$ & $4.94$ & $\sim 0 $ & $0.178$ & $\sim 0 $ \\
			FRB 20210603B & $122.846$ & $21.986$ & $636.1292$ & GRB 141121A & $122.667$ & $22.233$ & $0.298$ & $1.47$ & $\sim 0 $ & $0.245$ & $\sim 0 $ \\
			FRB 20211201B & $206.429$ & $43.997$ & $369.4879$ & GRB 080319A & $206.354$ & $44.083$ & $0.102$ & $2.0265$ & $\sim 0 $ & $0.215$ & $\sim 0 $ \\
			FRB 20220426D & $195.064$ & $32.127$ & $118.083$ & GRB 141220A & $195.050$ & $32.150$ & $0.026$ & $1.3195$ & $\sim 0 $ & $0.222$ & $\sim 0 $ \\
			FRB 20230508C & $324.113$ & $6.645$ & $471.4404$ & GRB 061110B & $323.900$ & $6.883$ & $0.318$ & $3.44$ & $\sim 0 $ & $0.234$ & $\sim 0 $ \\
			FRB 20230915C & $1.326$ & $31.878$ & $748.879$ & GRB 220101A & $1.379$ & $31.750$ & $0.136$ & $4.618$ & $\sim 0 $ & $0.222$ & $\sim 0 $ \\
			\hline
			\multicolumn{12}{l}{%
				\begin{minipage}{\linewidth}
					\smallskip
					\footnotesize \textbf{Note.} Within this table, SGRB 060502B is the only explicitly confirmed short gamma-ray burst. The associated FRBs in these pairs lack spectroscopically determined host galaxy redshifts. The redshift distribution used in the Bayesian analysis is inferred from their $\mathrm{DM_{E}}$ (see Section \ref{p_z}).
				\end{minipage}%
			} \\
		\end{tabular}
	\end{sidewaystable*}
	\clearpage

	\section{Discussion} \label{Discussion}
	
	\subsection{FRB 20180916B - AT2020hur}
	The spatial coincidence between the repeating FRB 20180916B and the optical transient AT 2020hur was first identified and analyzed by \citet{Li_2022}, who used a frequentist approach to derive a  significance level of 0.000418  for the assumption of $H_0$. Our  Bayesian inference for this pair performed solely using celestial positional information due to the lack of a direct redshift measurement for AT 2020hur yields an association probability of 0.9998. This result provides independent statistical verification of the association claim, and the consistency between these two distinct statistical frameworks (frequentist vs. Bayesian) strengthens the evidence for a genuine physical association between FRB 20180916B and AT 2020hur.
	Beyond the statistical cross verification between independent analytical frameworks, the two sources exhibit extremely tight observational spatial coincidence. They share the same host galaxy (the very system from which the reference redshift of AT 2020hur is adopted) \footnote{\url{https://www.rochesterastronomy.org/sn2020/index.html}}, with their measured positions consistent at the 1 arcsecond level, matching the localization uncertainty of this two sources. This direct observational evidence, combined with our Bayesian inference results, further reinforces the robustness of the physical association between this two sources.

	\subsection{FRB 20190309A - sGRB 060502B}
	For the candidate pair of FRB 20190309A and the short-duration gamma-ray burst (sGRB) 060502B, our full  Bayesian framework yields an association probability of 0.83, which is insufficient to claim a statistically significant physical association. This result appears less statistically definitive than the association claim presented in \citet{Lu_2024}, who adopted a frequentist approach and derived a significance level of 0.05 for the assumption of $H_0$. 
	We note that the association probability is coupled to the total size of the FRB and AT samples ($N_\mathrm{FRB}$ and $N_\mathrm{AT}$), as formalized in Equation \ref{P_asso} via the conservative prior adopted to mitigate the look-elsewhere effect. Specifically, our analysis is built on the second CHIME/FRB catalog, while the study of \citet{Lu_2024} was based on the first CHIME/FRB catalog. The second CHIME/FRB catalog has a sample size nearly an order of magnitude larger than the first catalog, leading to a stricter conservative prior and thus a more moderate association  probability, which is fully in line with our theoretical expectations. To further verify the consistency of our framework, we recalculated the association probability using our identical Bayesian formalism, with only the FRB sample size replaced by that of the first CHIME/FRB catalog. This recalculation yields a revised association probability of 0.93, consistent with the high significance reported in the previous study.

	\subsection{Other Potential Association Candidates}
	
	Beyond the two high significance pairs discussed above, we also evaluate additional candidate pairs proposed in previous literature.
	The FRB 20200405A-AT 2019wur pair was previously reported by \citet{Li_2022}, which has a very small angular separation of $0.00019$ deg ($\sim0.68$ arcsec). Given the lack of a direct redshift measurement for AT 2019wur, we performed a spatial-only Bayesian inference for this pair, consistent with our treatment of the FRB 20180916B-AT 2020hur pair. We derive an association probability of $ \sim 10^{-4}$, which is consistent with the low significance reported in \citet{Li_2022}. This low probability is driven primarily by the large positional uncertainty of FRB 20200405A, and we therefore rule out a physical association for this pair.
	
	We also assess the repeating FRB 20250316A, which has three potential transient counterparts (SN 2008X, SN 2009E, and AT 2025erx) all located within its host galaxy NGC 4141. For all three pairs, our Bayesian calculation yields association probabilities consistent with zero, ruling out any statistically significant physical association. This result is fully consistent with the precision localization analysis of FRB 20250316A from \citet{2025ApJ...989L..48C}, which found no coincident transient emission within the FRB's localization error region.
	
	Finally, we evaluate the FRB 20190412B-SN 2009gi pair, proposed as a potential association candidate by \citet{2025ApJ...992..127L}. While the authors retained this pair as a candidate based on independent physical constraints, they noted that its spatial overlap is consistent with a random chance alignment. Our Bayesian framework confirms the lack of statistically significant correlation for this pair, yielding an association probability of $ \sim 10^{-9}$. We note that although SN 2009gi has a well-measured spectroscopic redshift, the absence of a precise localization for FRB 20190412B precludes any definitive assessment of a potential physical association.
	
	Taken together, these cases highlight two key conclusions from our analysis. First, small angular separation alone is insufficient to confirm a physical association between an FRB and an astrophysical transient, even when both sources reside in the same host galaxy. Second, high-precision localization of FRBs is essential for a definitive assessment of potential associations. Our Bayesian framework naturally accounts for positional uncertainties in its inference: for FRBs with precise localization, it can reliably quantify the significance of physical associations, while for FRBs with large positional uncertainties, the derived association probability is significantly suppressed, correctly reflecting that a small angular separation provides no statistically meaningful information in this regime.
	
	These two empirical conclusions are not arbitrary. They are directly and intuitively captured by the mathematical structure of our posterior association probability $P_{\rm asso}$ (Equation \ref{P_asso}). As formalized in the Bayes factor (Equation \ref{eq:B}), the framework incorporates two critical terms that govern the association significance: the exponential factor $\exp(-\theta^2/(2\sigma^2))$, which quantifies the statistical penalty for angular separation $\theta$ relative to the combined positional uncertainty $\sigma$, and the geometric factor $1/(2\pi\sigma^2)$, which explicitly suppresses the association signal for sources with poor localization. This means that even for pairs with extremely small angular separation, a large $\sigma$ (i.e., low precision localization) will drastically reduce the Bayes factor and thus the posterior probability $P_{\rm asso}$, mathematically encoding the fact that small angular separation alone is statistically uninformative. Conversely, for FRBs with precise localization (small $\sigma$), the framework is highly sensitive to even moderate spatial coincidences, enabling robust quantification of association significance. In this way, the mathematical design of our formalism naturally reflects the two key observational lessons drawn from our case studies.

	\section{Conclusion} \label{conclusion}
	
The physical origin of FRBs remains unresolved, with the magnetar central engine model as the leading framework. A critical test of this model is the identification of physical associations between FRBs and the energetic astrophysical transients (ATs) that form magnetars, such as gamma-ray bursts (GRBs) and core-collapse supernovae. Previous association searches have been hampered by small FRB samples. In this work, we perform a systematic search for FRB-AT associations employing the largest available FRB dataset via a statistically rigorous 3D Bayesian inference framework.

Our analysis uses a cleaned sample of 3765 unique FRBs, combining the second CHIME/FRB catalog and 124 additional localized FRBs with measured redshifts. Our 3D Bayesian framework jointly incorporates angular separation, positional uncertainty, and self-consistent redshift probability distributions, with a conservative prior to suppress false positives from the look-elsewhere effect in large-scale catalog cross-matching.

We independently recover the previously reported high-significance association candidate between FRB 20180916B and AT 2020hur, with a spatial-only posterior association probability of 0.9998. For the second previously proposed candidate, FRB 20190309A and the short GRB 060502B, our full 3D framework yields a posterior association probability of 0.83, which is insufficient to claim a statistically significant physical association. Our updated results do not support the authenticity of this proposed association. Under our strict statistical framework, we find no new statistically significant FRB-AT associations, with all remaining candidates yielding association probabilities approaching zero.

The null detection of new associations in this large, statistically robust study suggests that genuine FRB-AT associations are far rarer than previously expected, with important implications for FRB progenitor models. Upcoming radio facilities with sub-arcsecond localization capabilities, such as the CHIME/FRB Outriggers and DSA-2000, will enable transformative advances in FRB-AT association searches, providing definitive clues to the physical origin of FRBs.

\section*{acknowledgments}
This work is supported by the National Natural Science Foundation of China (grant Nos.12494575,12203013).

	\bibliographystyle{aasjournal} 
	\bibliography{reference} 
	
\end{document}